\documentstyle[prl,aps,floats]{revtex}

\begin{document}
\draft

\title{First Measurement of $\pi^-{e}\rightarrow\pi^- e \gamma$\
       Pion Virtual Compton Scattering}

\author{
A.~Ocherashvili$^{12}$$^{,\P\P}$,
G.~Alkhazov$^{11}$,
A.G.~Atamantchouk$^{11}$,
M.Y.~Balatz$^{8}$$^{,\ast}$,
N.F.~Bondar$^{11}$,
P.S.~Cooper$^{5}$,
L.J.~Dauwe$^{17}$,
G.V.~Davidenko$^{8}$,
U.~Dersch$^{9}$$^{,\dag}$,
A.G.~Dolgolenko$^{8}$
% ...
\\                                                                            
\vskip 0.50cm                                                                 
\centerline{(SELEX Collaboration)}                                             
\vskip 0.50cm                                                                 
}                                                                             
\address{  
% ...
$^5$Fermilab, Batavia, IL 60510, U.S.A.\\
% ...
$^8$Institute of Theoretical and Experimental Physics, Moscow, Russia\\
$^9$Max-Planck-Institut f\"ur Kernphysik, 69117 Heidelberg, Germany\\
% ...
$^{11}$Petersburg Nuclear Physics Institute, St. Petersburg, Russia\\
$^{12}$Tel Aviv University, 69978 Ramat Aviv, Israel\\
% ...
$^{17}$University of Michigan-Flint, Flint, MI 48502, U.S.A.\\
% ...
}

\date{\today}
\maketitle
%
%
\begin{abstract}
Pion Virtual Compton Scattering (VCS) via the reaction
$\pi^-{e}\rightarrow\pi^-{e}\gamma$ was observed in the Fermilab E781
SELEX experiment. SELEX used a $600\,\mbox{GeV/c}$ $\pi^-$ beam
incident on target atomic electrons, detecting the incident $\pi^-$ and
the final state $\pi^-$, electron and $\gamma$. Theoretical
predictions based on chiral perturbation theory are incorporated into
a Monte Carlo simulation of the experiment and are compared to the data.
The number of reconstructed events (9) and their distribution with
respect to the kinematic variables (for the kinematic region studied)
are in reasonable accord with the predictions. The corresponding
$\pi^-$ VCS experimental cross section is 
$\sigma=38.8\pm{13}\,\mbox{nb}$, in
agreement with the theoretical expectation $\sigma=34.7\,\mbox{nb}$. 
\end{abstract}

\pacs{13.60.Fz,14.40.Aq}

\twocolumn
\input{psfig}
%
%
%
%
\section{Introduction}

The electric $(\bar{\alpha})$ and magnetic $(\bar{\beta})$ pion
polarizabilities characterize the pion's deformation in an electromagnetic
field, as occurs during $\gamma\pi$ Compton scattering. They depend on the
rigidity of the pion's internal structure as a composite particle, and are
therefore important dynamical quantities to test the validity of theoretical
models. Based on QCD chiral dynamics, the chiral perturbation theory
effective Lagrangian, using data from radiative pion
beta decay, predicts the pion electric and magnetic polarizabilities
$\bar{\alpha}_{\pi}$ = -$\bar{\beta}_{\pi}$ = 2.7 $\pm$ 0.4, expressed
in units of $10^{-43}\,\mbox{cm}^3$~\cite{Pennington,Pole,ho90}. Other
theoretical predictions are also available~\cite{Pennington}.  

\begin{figure}[h]
 \begin{center}
  \centerline{\psfig{figure=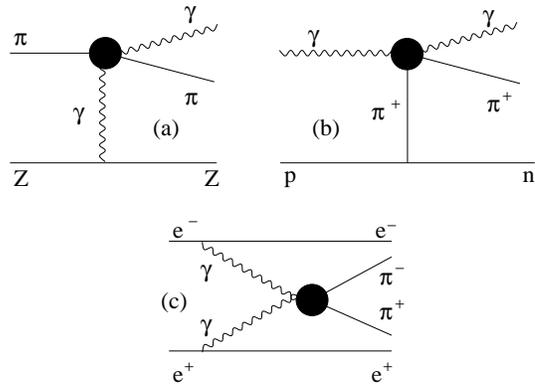,width=70mm}}
   \caption{Three pion Compton scattering reactions:
             (a) $\pi^- Z\rightarrow\pi^-{Z}\gamma$, 
             (b) $\gamma p\rightarrow\gamma\pi^+ n$,
             (c) $e^+ e^-\rightarrow{e^+ e^-}\pi^-\pi^+$.}   
 \label{fig.polpos}
 \end{center}              
\end{figure}

The pion polarizabilities are usually investigated via their
effect on the shape of the measured $\gamma\pi\rightarrow\gamma\pi$
Real Compton Scattering (RCS) angular distribution, as in Ref.~\cite{kbp}.
Since pion targets are unavailable, pion RCS is approximated
using different artifices, as shown in Fig.~\ref{fig.polpos}: the 
$\pi^- Z\rightarrow\pi^-{Z}\gamma$ Primakoff~\cite{Antipov} and 
$\gamma p\rightarrow\gamma\pi^+ n$ radiative pion photoproduction reactions
\cite{aibergenov}; or by the crossing symmetry
$\gamma\gamma\rightarrow\pi^+ \pi^-$ two-photon reaction~\cite{Pole,Boyer}.
In the Primakoff scattering, a high energy
pion scatters from a (virtual, practically real)  photon in the Coulomb field
of the target nucleus. Values of $\bar{\alpha}$
measured by these experiments are given in Table~\ref{tab:mespol}. 
\begin{table}[h]
\begin{center}
\begin{tabular}{|c|c|c|}
\hline
Reaction& $\bar{\alpha}\,\,[10^{-43}\,\mbox{cm}^3]$ & Reference \\
\hline
$\pi^-Z\rightarrow \pi^-Z\gamma$ &$6.8\pm 1.4\pm 1.2$ &~\cite{Antipov} \\
$\gamma p \rightarrow \gamma \pi^+ n$ &$20\pm 12$     &~\cite{aibergenov}\\
$\gamma\gamma \rightarrow \pi^+\pi^-$ &$2.2\pm 1.6$   &~\cite{Pole,Boyer}\\
\hline
\end{tabular}
\caption{Experimental values of $\bar{\alpha}$.}
\label{tab:mespol}
\end{center}
\end{table} 
They cover a large range of values and have large uncertainties. New
high precision pion polarizability measurements are therefore needed. 
Electromagnetic studies with virtual photons have the advantage that the
energy and three-momentum of the virtual photon can be varied
independently. In the pion $\gamma^*\pi\rightarrow\gamma\pi$ VCS reaction,
where the initial state photon is virtual (far from the quasi-real
photons of Primakoff scattering) and the final state photon is
real, polarizabilities can be measured in the space like region,
inaccessible by RCS~\cite{Scherer_I}. We thereby access the
so-called electric $\bar{\alpha}(q^2)$ and magnetic $\bar{\beta}(q^2)$
generalized polarizabilities of the pion~\cite{newalpha2},
where $\bar{\alpha}(0)$ and $\bar{\beta}(0)$ correspond to the RCS
$\bar{\alpha}$ and $\bar{\beta}$ pion Compton polarizabilities. The
$q^2$-dependent $\bar{\alpha}(q^2)$ determines the change
$\Delta{F}(q^2)$ in the pion charge form factor $F(q^2)$ in the presence
of a strong electric field. The Fourier transform of $\bar{\alpha}(q^2)$
provides a picture of the local induced pion charge polarization density
\cite{pionffc}.  Similarly, first experiments~\cite{dhos} and
calculation~\cite{pas} have been carried out for proton VCS via $e p
\rightarrow e p \gamma$.  

We study experimentally the feasibility of extracting the ``pion VCS''
reaction:  
\begin{equation}
\label{react}
\pi^- e\rightarrow\pi^- e \gamma,
\end{equation}
as a step in developing pion VCS as a new experimental tool for pion
polarizability measurements. The data were taken with the Fermilab
E781 SELEX spectrometer~\cite{prop}. We used a $600\,\mbox{GeV/c}$
$\pi^-$ beam incident on target atomic electrons, detecting the
incident $\pi^-$ and the final state $\pi^-$, electron and
$\gamma$. Theoretical predictions~\cite{pvcs,scherer} based on chiral
perturbation theory are incorporated into a Monte Carlo simulation of
the experiment and are compared to the data. With this theory, we
calculated the total cross section (as described later) of the
Eq. (\ref{react}) process for a limited kinematic region discussed
later [Eq. (\ref{s1s2r})], using the Monte Carlo integration program
VEGAS~\cite{vegas}. The result is $\sigma(total)=34.7\pm{0.1}\,\mbox{nb}$. 
For the kinematic range considered, the integrated cross sections are
not sensitive to the polarizabilities. We nonetheless chose this
region  in order to obtain sufficient statistics for a first study of
the reaction.   
\section{VCS kinematics and theoretical differential cross section}
\label{sec:vcsk}
\begin{figure}[h]
  \begin{center}
    \leavevmode
   \centerline{\psfig{figure=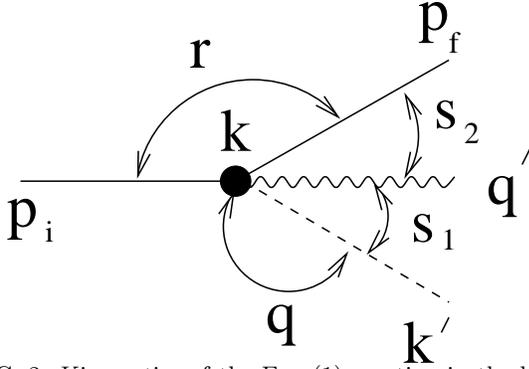,width=70mm}}
    \caption{Kinematics of the Eq. (\ref{react}) reaction in the laboratory
      frame. The incoming pion beam with 4-momentum
      $p_i$ interacts with the target electron with
      4-momentum $k$ and produces the outgoing pion with
      4-momentum $p_f$, $\gamma$ with 4-momentum $q'$,
      and electron with 4-momentum $k'$.}
    \label{fig.kin}
  \end{center}
\end{figure}

We study reaction (\ref{react}) in terms
of the following five independent invariant variables:
\begin{eqnarray}
\label{ivar}
s=(p_i+k)^2, \nonumber \\
s_1=(k'+q')^2, &
s_2=(p_f+q')^2, \\
r^2=(p_i-p_f)^2, &
q^2=(k'-k)^2. \nonumber
\end{eqnarray}
Here $p_i$ is the 4-momentum of the incoming pion, $k$ is the
4-momentum of the target electron, and $p_f$, $k'$, $q'$ are the
4-momenta of the outgoing pion, electron, and photon, respectively, as
shown in Fig.~\ref{fig.kin}. 
The differential cross section of reaction (\ref{react}) in the convention of
Ref.~\cite{bdrel} reads as:  
\begin{eqnarray}
  \label{bdcross}
d \sigma=\frac{m_e^2}{8E_fE_{k'}E_{q'}}
\frac{1}{\sqrt{(p_i\cdot
    k)^2-M_{\pi}^2m_e^2}}\frac{1}{(2\pi)^5}\cdot \nonumber\\
|\overline{{\cal M}} 
|^2 \delta^4(p_i+k-p_f-k'-q')d^3p_fd^3k'd^3q'. 
\end{eqnarray}
The invariant amplitude $\cal{M}$ contains the complete information on
the dynamics of the process. The quantity $|\overline{\cal{M}}|^2$
indicates the sum over the final states and the average over the
initial spin states. The fourfold differential cross section in terms
of independent invariant variables of Eq. (\ref{ivar}) involves a
kinematical function $\lambda$~\cite{kinematic} and a Jacobian matrix
$\Delta_4$ defining the phase space of the physical areas, and is given by:
\begin{equation}
\label{eq:xsection}
\frac{d\sigma}{ds_1ds_2dq^2dr^2}=
\frac{1}{{(2\pi)}^5}
\frac{2m_e^2}{\lambda{(s,m_e^2,m_{\pi}^2)}}
\frac{\pi}{16}
\frac{1}{{(-\Delta_4)}^{1/2}}
{|\bar{\cal M}|}^2.
\end{equation}
Since the variable $s$, involving the energies of the beam pion and of
the target electron, is fixed, differential cross section
(\ref{eq:xsection}) actually depends on four variables.  

\begin{figure}[h]
  \begin{center}
   \leavevmode
    \centerline{\psfig{figure=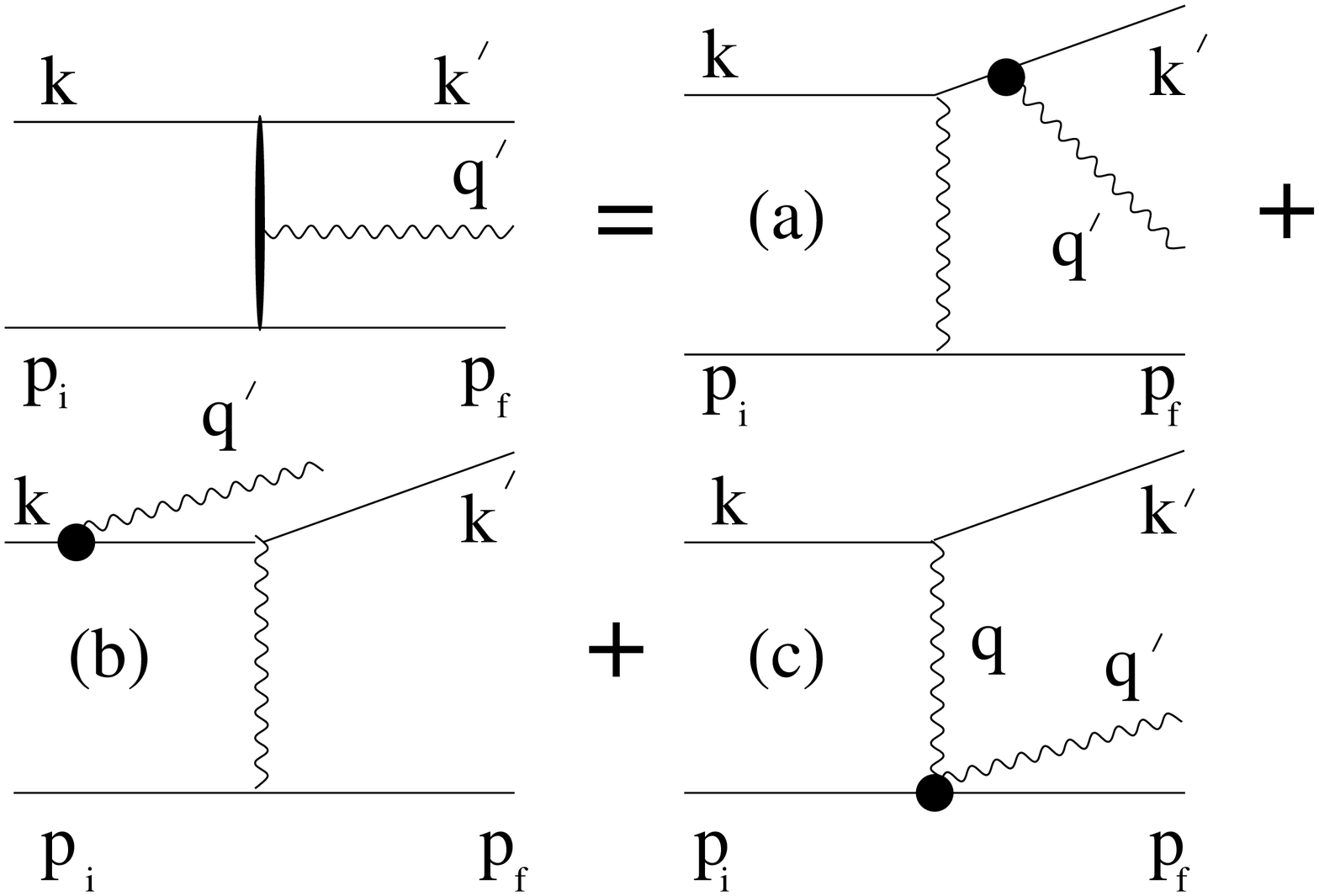,width=70mm}}
    \caption{The three Feynman diagrams corresponding to the 
             $\pi^-{e}\rightarrow\pi^-{e}\gamma$. In the one
      photon exchange approximation, (a) and (b) correspond to the BH
      process, while (c) corresponds to the VCS process. Here, $q$ is
      the 4-momentum of the virtual photon, $q=k'-k$.} 
    \label{fig.fdiag}
  \end{center}
\end{figure}

In reaction (\ref{react}), the final photon can be emitted either by the
electron or by the pion, as shown in Fig.~\ref{fig.fdiag}. The first
process is described by the Bethe-Heitler (BH) amplitude
(Fig.~\ref{fig.fdiag}a, b), calculable from quantum electro-dynamics.
The second process is described by the VCS (Fig.~\ref{fig.fdiag}c)
amplitude. Since the source of the final photon emission is
indistinguishable, one obtains the following form of the matrix
element of reaction (\ref{react})~\cite{pvcs}:
\begin{eqnarray}
  \label{mbhmvcs}
|\overline{\cal M}|^2=\frac{1}{2}\sum
({\cal M}^{VCS}+{\cal M}^{BH})({\cal M}^{VCS*}+{\cal M}^{BH*})=\nonumber\\
|\overline{{\cal M}^{BH}}|^2+|\overline{{\cal M}^{VCS}}|^2+\nonumber\\
|\overline{{\cal M}^{VCS}{\cal M}^{BH*}+{\cal M}^{VCS*}{\cal M}^{BH}}|,
\end{eqnarray}
where ${\cal M}^{BH}$ and ${\cal M}^{VCS}$ are amplitudes of the BH
and VCS processes.

The general features of the four-fold differential cross section
can be inferred from Eq. (\ref{eq:xsection}) and matrix element
calculations. The $s_1$-dependence is dominated by the
$(s_1-m^2_e)^{-1}$ pole of BH, the cross section varies accordingly, and is
only slightly modified by the $s_1$ dependence of the VCS
amplitude. The $s_2$ dependence is dominated by the
$(s_2-M^2_{\pi})^{-1}$ pole of the VCS amplitude with modification of
$s_2$ dependence of the BH amplitude, but in this case the
modification is not as small as in the case of $s_1$. The $r^2$ 
dependence is determined by the $r^{-4}$ pole of the BH amplitude, and
the $q^2$ dependence follows the $q^{-4}$ pole behavior of typical
electron scattering. The energy of the outgoing pion is expected to be high
while the angle is expected to be small according to the $r^{-4}$
behavior of the cross section. The energy of the outgoing photon is
mainly expected to be low, as follows from the infrared divergence
of the BH amplitude. The angular behavior of the outgoing photon is
determined by the $(s_1-m^2_e)^{-1}$ and $(s_2-M^2_{\pi})^{-1}$ poles
of the BH and VCS amplitudes. The more interesting photons related to
the generalized polarizabilities are expected to have higher
energies. The behavior of the outgoing electron is completely
determined by the $q^{-4}$ behavior of the cross section. 
\section{Experimental apparatus and trigger}
Our data were taken with the Hadron-Electron (HE) scattering trigger 
\cite{trigger} of experiment E781/SELEX at Fermilab. SELEX uses a
negatively charged beam of $600\,\mbox{GeV/c}$ with full width momentum spread
of dp/p=$\pm$ 8\%, and an opening solid angle of 0.5 $\mu$sr. The beam
consisted of approximately 50\% $\pi^-$ and 50\% $\Sigma^-$. SELEX
used Copper and Carbon targets, totaling 4.2\% of an interaction
length, with target electron thicknesses of $0.676\,\mbox{barn}^{-1}$ and
$0.645\,\mbox{barn}^{-1}$, respectively.   

The experiment focused on charm baryon hadroproduction and
spectroscopy at large-$x_F$. The spectrometer hosted several projects
which exploit physics with a small number of tracks compared to
charm. SELEX had good efficiency for detecting all particles in the
final state since the produced particles and decay fragments at
large-$x_F$ production are focused in a forward cone in the laboratory
frame. Other requirements in the charm program for background
suppression include good vertex resolution and particle identification
over a large momentum range.      

Four dipole magnets divide SELEX into independent spectrometers (Beam,
M1, M2 and M3) each dedicated to one special momentum region. Each
spectrometer had a combination of tracking detectors. The
M1, M2 and M3 spectrometers included electromagnetic 
calorimeters. The $\pi-e$ separation for 
hadron-electron scattering used the M2 particle identification
transition radiation detector and also the electromagnetic
calorimeter.  

The HE scattering trigger was specialized for separation of
hadron-electron elastic scattering events. The trigger 
used information from a charged particle detectors just downstream of
the target and, from a hodoscope just downstream of M2, to determine charged
particle multiplicity and charge polarity. For the HE requirement, no
electromagnetic calorimeter information was included. Therefore, the
data collected via this trigger include hadron-electron elastic and
inelastic scattering events. 
\section{Monte Carlo Simulation}

\begin{figure}[h]
  \begin{center}
    \leavevmode
    \centerline{\psfig{figure=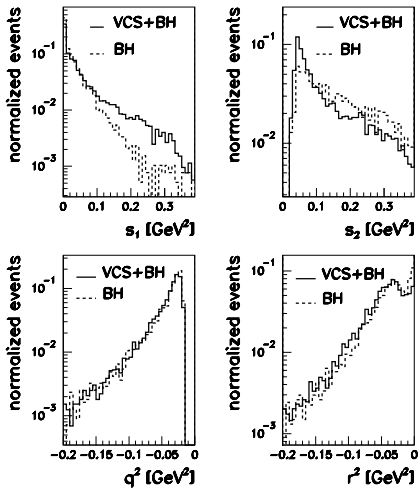,width=70mm}}
    \caption{Generated distributions of events plotted with
      respect to the Mandelstam invariants(solid line corresponds to
      the generation with full VCS amplitude and dashed line
      corresponds to the generation where only BH amplitude was used.} 
    \label{fig.geninv}
  \end{center}
\end{figure}

Monte Carlo simulations were carried out for $\pi^-$ VCS signal (\ref{react})
and background event distributions with respect to the four invariants
s$_1$, s$_2$, q$^2$, r$^2$. We used the SELEX GEANT package GE781~\cite{mc2}.
The Monte Carlo (MC) study was carried out in four steps:  
\begin{enumerate}
\item A VCS event generator was written to search for the regions of
      phase space where the data are sensitive to pion structure. 
      Several event generators were made to simulate a variety of
      expected backgrounds.   
\item The event generators were implemented into the simulation
      package. We studied the resolution, detection 
      efficiency, geometric and trigger acceptances for the signal and
      background.  
\item The offline analysis procedure was developed and tuned to devise
      software cuts eliminating background while preserving a VCS signal.
\item Finally we estimated the expected number of $\pi^-$ VCS events. 
\end{enumerate}
The VCS event generator is written, based on differential
cross section (\ref{eq:xsection}), matrix element calculations, and
3-body final state kinematics. The acceptance-rejection method
\cite{ARM} is used for event generation. The VCS cross section
increases rapidly when the direction of the outgoing real photon is
close to the direction of one of the outgoing charged particles (due
to the $(s_1-m^2_e)^{-1}$ pole of BH and the $(s_2-M^2_{\pi})^{-1}$
pole of VCS), or when the energy of the outgoing real photon comes
close to zero (due to the infrared divergence of the BH
($(s_1-m^2_e-q^2+r^2)^{-1}$ pole). Therefore, if events are generated 
in the pole region, then the efficiency of the acceptance-rejection
method can be rather low, for the more interesting non-pole regions. In order 
to generate events at an acceptable rate, the pole regions are
eliminated. Invariants are generated in the following regions:  
\begin{eqnarray}
\label{s1s2r}
1000m^2_e\leq{s_1}\leq{M^2_{\rho}}, \nonumber \\
2M^2_{\pi}\leq{s_2}\leq{M^2_{\rho}}, \nonumber \\
-0.2\;GeV^2<q^2<-0.032\;GeV^2, \nonumber \\
-0.2\;GeV^2<r^2<-2m_e E_{\gamma}(min)+q^2+s_1-m^2_e.
\end{eqnarray}
For the photon minimum energy, we choose $E_{\gamma}(min)=5$ GeV to
cut the infrared divergence of BH, and to be above the calorimeter
noise. Since the VCS calculation does not explicitly include the
$\rho$ resonance, we choose upper limits of $M^2_{\rho}$ for $s_1$ and
$s_2$.  In Fig.~\ref{fig.geninv}, we show the generated distribution
of events plotted with respect to the Mandelstam invariants, without
correction for any acceptances. The solid and dashed curves are for
BH+VCS and BH respectively. The VCS amplitude clearly affects the shape
of these distributions for s$_1$ and s$_2$. For q$^2$ and r$^2$ the
effect is difficult to see for the statistics shown. Taken together,
the experiment therefore is potentially sensitive to the pion VCS amplitude.  
Since $\pi{/e}$ separation is not 100\% efficient, all interactions
which produce two negatively charged tracks and a photon in the final
state can generate the pattern of the pion VCS; i.e., can create
background to the required measurement. For the background simulation,
as well as in case of the VCS simulation, we require 5~GeV minimum
energy for the photon.    

\begin{figure}[h]
  \begin{center}
    \leavevmode
    \centerline{\psfig{figure=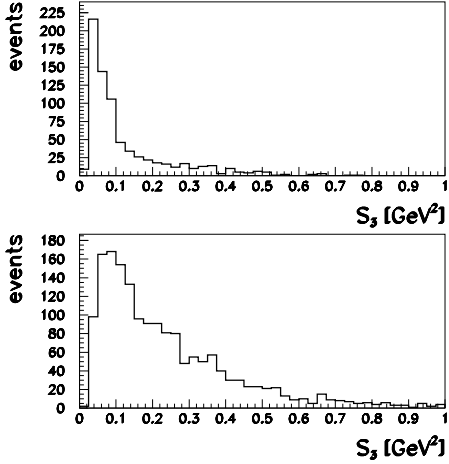,width=70mm}}
    \caption{Background via meson production: simulated $s_3$
             distribution for $\pi^- e\rightarrow\rho e'$
             (upper), and for $\pi^- A(Z)\rightarrow{M} A(Z)$
             (lower), with arbitrary normalization.} 
    \label{fig.s3mesontalk}
  \end{center}
\end{figure}

The dominant background process for pion VCS is $\pi^-{e}$
elastic scattering followed by final state interactions of the outgoing
charged particles, such as Bremsstrahlung. The MC simulation
shows that 28\% of the original  $\pi^-{e}$ elastic scattering events
generate more than 5 GeV Bremsstrahlung photons somewhere in the SELEX
apparatus. We therefore consider the $s_3$ invariant mass of the outgoing
$\pi^-{e}$ system:   
\begin{equation}
\label{s3dec}
s_3=(p_f+k')^2.  
\end{equation}
We expect $s_3=s$ for elastic scattering, and
$s_3=s_1-s_2+m_e^2+M^2_{\pi}$ for VCS.   
\begin{figure}[h]
  \begin{center}
    \leavevmode
    \centerline{\psfig{figure=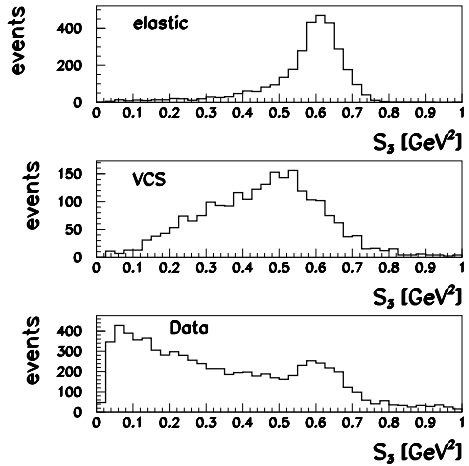,width=70mm}}
    \caption{Comparison of simulated $s_3$ distributions for $\pi^-{e}$
             elastic scattering (top) and pion VCS (middle) with the
             data (bottom).}
    \label{fig.s3talk}
  \end{center}
\end{figure}
Fig.~\ref{fig.s3talk} shows generated and reconstructed elastic and
VCS events subjected to the same charged particle track reconstruction
and trigger requirements as the data. 
The data shown are only  
in the kinematic region of
Eq.~\ref{s1s2r}.
As seen from Fig.~\ref{fig.s3talk}, 
the events in the data 
at low $s_3$ do not arise from $\pi^- e$ elastic or $\pi^-$
VCS. To describe the events with low $s_3$, we simulate backgrounds
(with corrections for acceptances) from the interactions:  
\begin{eqnarray}
\label{rhoe}
\pi^-~e\rightarrow{M}~e,
\end{eqnarray}
\begin{eqnarray}
\label{meson}
\pi^-~A(Z)\rightarrow{M}~A(Z),
\end{eqnarray}
where $M$ in Eqs. (\ref{rhoe}) and (\ref{meson}) is an intermediate
meson state which can decay via $\pi^-\pi^0$, $\pi^-\eta$, $\pi^-\omega$,
etc. Considering the threshold energies of reaction (\ref{rhoe}), only
$\rho$ meson production is allowed at SELEX energies. 
Fig.~\ref{fig.s3mesontalk} shows the simulated $s_3$ distributions for
reactions (\ref{rhoe}-\ref{meson}), with the same acceptance
requirements as in Fig.~\ref{fig.s3talk}.   

SELEX GE781 allows us to estimate the background rate from $\pi^- e$
elastic scattering. However, it is very difficult to estimate the
background rate from all neutral meson ($\pi^0$, etc.) production
reactions, such as those shown in Fig.~\ref{fig.s3mesontalk}. The data
of Fig.~\ref{fig.s3talk}(bottom) are 
qualitatively well described by a combination of elastic
(Fig.~\ref{fig.s3talk}~top) and 
meson production (Fig.~\ref{fig.s3mesontalk}~bottom) channels. The
contributions of VCS (Fig.~\ref{fig.s3talk}~middle) 
and $\rho$ production (Fig.~\ref{fig.s3mesontalk}~top) are relatively
much lower. We do not show the quantitative sum of all contributions,
because of the difficulty to estimate the absolute yields of all meson
production channels. Instead, we seek a set of cuts which remove as
``completely'' as possible the background from all $\gamma$ sources,
arising from neutral meson decays. 

Since we measure all outgoing particles, the reaction kinematics are
overdetermined. Therefore, in the data analysis, a constrained $\chi^2$
fitting procedure~\cite{brandt} is used. A veto condition is used
based on a 2-body final state constrained $\chi^2$ kinematic fit for reduction
of the background from $\pi^-{e}$ elastic scattering~\cite{mpi}. A
3-body final state constrained $\chi^2$ kinematic fit is used for extracting
the pion VCS signal.   

A final state electron can arise from photon conversion.
For this background, the electron  is not
created at
the same vertex as the pion. Consequently, the
quality of the 3-track vertex reconstruction should be low. Also, the 
simulation shows that no VCS outgoing particles hit the upstream
photon calorimeter. Therefore, in addition to the constrained
kinematic fitting cuts, we use restrictions on the vertex quality and 
on the total energy deposit in the upstream photon calorimeter to
suppress backgrounds. 

\begin{figure}[h]
  \begin{center}
  \centerline{\psfig{figure=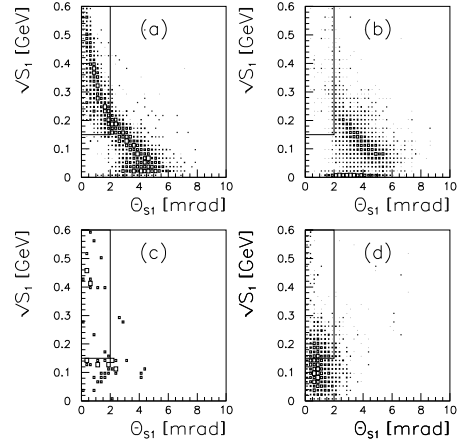,width=70mm}}
    \caption{$\sqrt{s_1}$, $\Theta_{s_1}$ correlation for (a)
      VCS, (b) $\pi^-{e}$ elastic scattering, (c) intermediate meson, and (d)
      $\rho$ meson production. Only events inside the
      $\Theta_{s_1}$-$\sqrt{s_1}$ region indicated by the box were
      accepted for the further analysis.}
    \label{fig.s1ths1}
  \end{center}
\end{figure}

\begin{figure}[h]
  \begin{center}
  \centerline{\psfig{figure=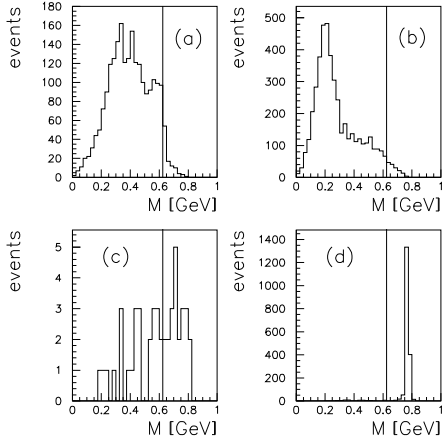,width=70mm}}
    \caption{Generated distribution of the $M$ invariant for: (a) VCS,
      (b) elastic scattering, (c) meson production reactions
      (\ref{meson}), and (d) $\rho$ production reaction
      (\ref{rhoe}). The vertical lines show the 0.625 GeV M-cut position.}
    \label{fig.mass}
  \end{center}
\end{figure}

An additional way to reduce the background from $\pi^-{e}$ elastic
scattering is to use the $\gamma{e}$ invariant mass $s_1$ and the
$\Theta_{s_1}$ angle (angle between $p_i$ and $(q'+k')$; see
Fig.~\ref{fig.kin}). The simulation shows that if the final photon 
is emitted via electron bremsstrahlung, the value of $s_1$ should be
low. On the other hand if the photon and electron are produced via $\pi^0$, 
$\eta$, or $\omega$ meson, $s_1$ will ``remember'' the mass of the
parent particle. Holding the value of the $s_1$ invariant to be between the
squared mass of $\pi^0$ and  $\eta$ mesons cuts the background from the
electron bremsstrahlung and from reactions (\ref{meson}). Fig.~\ref{fig.s1ths1}
shows the distributions of VCS and background simulations. It is seen
that the region
with higher $\sqrt{s_1}$ and lower $\Theta_{s_1}$ are mostly populated
with pion VCS. 
Another possibility to reduce the background from interactions
(\ref{rhoe}-\ref{meson}) is to cut on the invariant $M$, defined as:
\begin{equation}
  \label{mass}
  M=\sqrt{(P_{\pi}+P_e-P_{e'})^2}.
\end{equation}
For elastic events $M=M_{\pi}$, for VCS $M=\sqrt{s_2}$, for $\rho$
production $M=M_{\rho}$ (see Fig.~\ref{fig.mass}).   
 The set of the final cuts follows: 
\begin{itemize}
\item {\bf 1}, fulfillment of the pion VCS pattern with:
  identified electron; E$_{\gamma}\ge$5 GeV for photons observed at
  laboratory angles less than 20~mrad in the downstream
  electromagnetic calorimeters; no additional tracks and vertices;
  the total energy deposit is less than 1~GeV in the upstream
  electromagnetic calorimeter, covering detection angles greater than
  30~mrad. 
\item {\bf 2}, Eq. (\ref{s1s2r}) ranges for the invariants, with 
$s_1$(min)=0.0225 GeV$^2$.
\item {\bf 3}, $\chi^2_{elastic}>20$, following a constrained fit
  procedure~\cite{brandt,mpi}.
\item {\bf 4}, $\chi^2_{VCS}\le5$, following a constrained fit
  procedure~\cite{brandt,pvcs}.
\item {\bf 5}, $\Theta_{s_1}<$2 mrad, $M\le0.625$ GeV.
\end{itemize}
To estimate the number of expected VCS events, as well as the yields of
other $\pi^-{e}$ elastic or inelastic scattering events, we use:
\begin{equation}
\label{piel}
N_{\pi{e}}=N_{\pi}\cdot\sigma\cdot{N_T}\cdot\epsilon_{ex}\cdot\epsilon_r.
\end{equation}
Here $N_{\pi{e}}$ is the number of events observed for a particular
$\pi{e}$ reaction, $N_{\pi}$ is the number of incident beam pions
($\sim 4.4\cdot 10^{10}$ as measured by beam scalers and including
particle identification), $\sigma$ is the cross section for the
particular reaction, 
$\epsilon_r$ is the offline reconstruction efficiency (36.6\% for
elastic, 2.65\% for VCS) of the studied process, and 
$N_T$ is the target electron density. Since not all experimental
properties are implemented in GEANT, an additional efficiency factor
$\epsilon_{ex}$ is included in Eq.~\ref{piel}. This efficiency factor
is common to $\pi{e}$ elastic and pion VCS reactions. The value
$\epsilon_{ex}$ is calculated by comparison of the actual number of observed
$\pi e$ elastic scattering events to the expected number of events.
The common efficiency arises since these two reactions
have practically the same $q^2$ dependence; $q^2$ being
the only kinematical parameter relevant for the trigger and
first order data reduction procedure. We calculate the experimental efficiency
$\epsilon_{ex}$ (13.4\%) from $\pi{e}$ elastic scattering, as described
in Ref.~\cite{pvcs}. For extraction of the reference $\pi{e}$ elastic
scattering events, we employ the cuts of the SELEX $\pi{e}$ elastic
scattering analysis~\cite{mpi}. The cuts described above, considering
the studied sources of background, improve the signal/noise ratio
from less than 1/400 to more than 361/1. For these estimates, we used
the following cross sections: $\sigma_{VCS}=34.7\,\mbox{nbarn}$ and
$\sigma_{elastic}=4.27\,\mu\mbox{barn}$ for 
$\pi e$ scattering; $\sigma_{Primakoff}(C\;target)=0.025$ mbarn and
$\sigma_{Primakoff}(Cu\;target)=0.83\,\mbox{mbarn}$ for the $Z^2$-dependent
Primakoff scattering. Based on the calculated pion VCS cross section,
we expect $\sim$8 events in the $\pi^-$ data sample.

The effect of the above enumerated cuts on the VCS signal, and on the
backgrounds coming from $\pi{e}$ elastic scattering and Primakoff
meson production, are listed in Table~\ref{tab:mc_cut_acc}. The
results are based on the estimated relative cross section for these
three processes~\cite{pvcs}. The cuts are very effective in removing
backgrounds while retaining signal events, such that the final event
sample is essentially pure pion VCS. 

\begin{table}[h]
\begin{center}
\begin{tabular}{|c|c|c|c|}
\hline
Cuts&VCS&Elastic&Meson production\\
\hline
1&29.8 &32.1  &0.24  \\
2& 9.97& 9.21 &0.03  \\
3& 8.89& 0.03 &0.03  \\
4& 3.58& 0.003&0.004 \\ 
5& 2.56& $<0.001$&$<3. \times 10^{-6}$\\
\hline
\end{tabular}
\caption{
Percentages of the remaining events after using the cuts for the MC
simulated $\pi$ VCS and
background events.} 
\label{tab:mc_cut_acc}
\end{center}
\end{table}
\section{Data Analysis}
\begin{figure}[h]
  \begin{center}
  \centerline{\psfig{figure=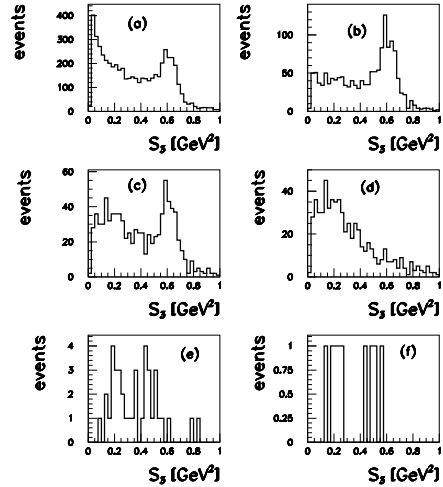,width=70mm}}
    \caption{Effect of the cuts on $s_3$ invariant, (a) raw
      distribution, (b) after cut 1, (c) after cut 2, (d) after cut 3,
      (e) after cut 4, (f) final distribution of $s_3$ invariant.}
    \label{fig.effs3data}
  \end{center}
\end{figure}

In the first stage analysis, events containing one identified electron are
selected. For these events, particle trajectories are checked if they
form a vertex inside the target material. An event is accepted if it
contains exactly three tracks, including the beam particle and an electron
candidate, and forms one vertex in the target. 
We consider data only  
in the kinematic region of
Eq.~\ref{s1s2r}.
On the accepted data
set, we applied the system of the cuts discussed above.
The working statistics with these cuts on the data are given in 
Table~\ref{tab:fincutsrej} and in Fig~\ref{fig.effs3data}, where we show
the effects
of the cuts on the $s_3$ invariant.
The effect of the cuts on the data (Table~\ref{tab:fincutsrej}) is comparable
to the effect on the simulated VCS events (Table~\ref{tab:mc_cut_acc}). 
\begin{table}[h]
\begin{center}
\begin{tabular}{|c|c|}
\hline
cuts&\% of remain events \\
\hline
1&26.8\\
2&13.9\\
3& 9.93\\
4& 0.61\\
5& 0.13\\
\hline
\end{tabular}
\caption{Percentages of events remaining after using the cuts.}
\label{tab:fincutsrej}
\end{center}
\end{table}
Finally 9 events (with a statistical uncertainty $\pm3$) were accepted
as pion VCS. 
The corresponding $\pi$ VCS experimental cross section 
based on Eq. (\ref{piel}) under the assumption that the background has
been completely eliminated is $\sigma=38.8\pm{13}\,\mbox{nb}$, in agreement
with the theoretical expectation $\sigma=34.7\,\mbox{nb}$. The error given
is only statistical, and does not include possible systematic
uncertainties in the efficiency product $\epsilon_{ex}\cdot\epsilon_r$
in Eq. (\ref{piel}). 

\begin{figure}[h]
  \begin{center}
   \centerline{\psfig{figure=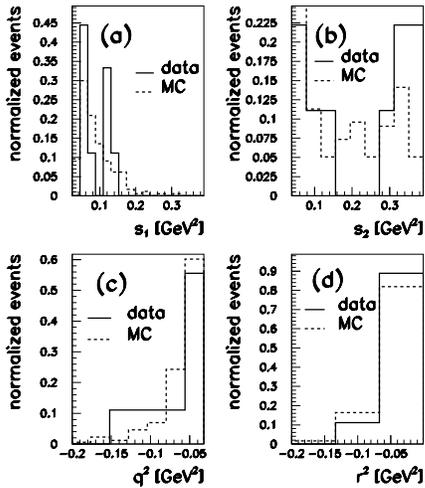,width=70mm}}
    \caption{Comparison of data and theory normalized distributions
             with respect to: (a) $s_1$, (b) $s_2$, (c) $q^2$, and (d)
             $r^2$. The solid line corresponds to data, the dashed
             line corresponds to simulations with $\bar{\alpha}=2.7$.
             The simulation with $\bar{\alpha}=6.8$ gives practically
             the same result.}
    \label{fig.invpim}
  \end{center}
\end{figure}

The comparisons of reconstructed (data) and generated (theory) event
distributions, normalized to unit area are shown in
Fig.~\ref{fig.invpim} with respect to the four invariants
$s_1,\;s_2,\;q^2$ and $r^2$ of Eq. (\ref{ivar}), shown with binning
that matches the experimental resolutions. 
The resolution
for each variable was found by Monte Carlo simulation, comparing
reconstructed and generated events.  
To check whether the data and theory (MC) distributions are consistent
with each other, we use the K-S test of Kolmogorov and
Smirnov~\cite{ks}. The K-S test is based on normalized cumulative
distribution functions (CDF). We use the K-S $D$ statistic, as a measure 
of the overall difference between the two CDFs. It is defined as the
maximum value of the absolute difference between two CDFs. The
significance level for a value $D$ (as a disproof of the null
hypothesis that the distributions are the same) is given by the
probability $P(D)$~\cite{ks}. A high value of $P(D)$ means that the
data and theory CDF are consistent with one another. 

\begin{figure}[h]
  \begin{center}
    \centerline{\psfig{figure=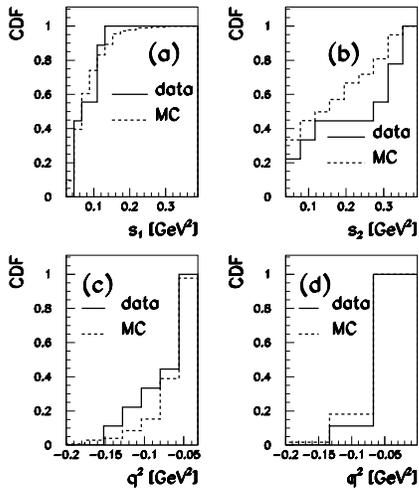,width=70mm}}
    \caption{Normalized cumulative distributions for data and theory versus:
             (a) $s_1$, (b) $s_2$, (c) $q^2$, and (d) $r^2$ invariants.
             The solid line corresponds to data, the dashed line
             corresponds to MC theory simulations with 
             $\bar{\alpha}=2.7$.The simulation with $\bar{\alpha}=6.8$
             gives practically the same result.}
    \label{fig.discomprob}
  \end{center}
\end{figure}

\begin{figure}[h]
  \begin{center}
   \centerline{\psfig{figure=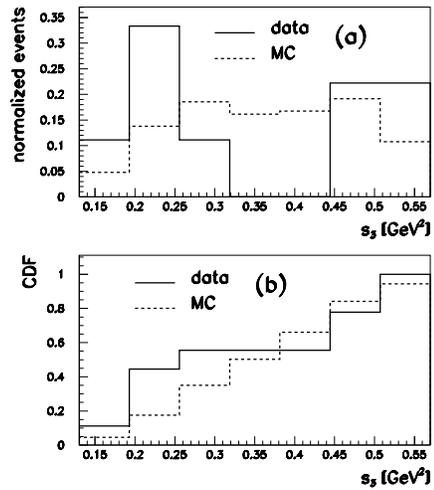,width=70mm}}
    \caption{Comparison of data and theory normalized distributions
             with respect to $s_3$ (a), and normalized cumulative
             distributions for data and theory versus $s_3$ (b).The
             solid line corresponds to data, the dashed line 
             corresponds to MC theory simulations with 
             $\bar{\alpha}=2.7$.The simulation with $\bar{\alpha}=6.8$
             gives practically the same result.}
    \label{fig.s3compks}
  \end{center}
\end{figure}

Following the K-S procedure, we calculate the normalized cumulative
distributions of data and theory, corresponding to
Figs.~\ref{fig.invpim}a-d.The K-S $D$ statistic, and the K-S
probabilities for consistency of theory/data distributions, are given
in Table~\ref{tab:ks} and Figs.~\ref{fig.discomprob}-~\ref{fig.s3compks}.  
\begin{table}[h]
\begin{center}
\begin{tabular}{|c|c|c|}                           
\hline
variable&$D$&$P(D)$\\
\hline
$s_1$&0.18&0.90\\
$s_2$&0.27&0.83\\
$s_3$&0.27&0.92\\
$q^2$&0.18&0.99\\
$r^2$&0.07&0.99\\
\hline
\end{tabular}
\caption {Value of K-S $D$ statistic, and probabilities $P$, for
  comparison of data with theory for $\bar{\alpha}=2.7$. The
  comparison of data with theory for $\bar{\alpha}=6.8$ gives
  practically the same result.}
\label{tab:ks}
\end{center}
\end{table}
The experimental and theoretical CDFs for s$_1$ and r$^2$ look
similar. For s$_2$ and q$^2$, some regions of q$^2$ (s$_2$) have the
experimental CDFs larger (smaller) than the theoretical CDFs. The
values of P(D) (see Table~\ref{tab:ks}) are sufficiently high for all
five CDF's, as expected if the experimental data are consistent with
theory.
In addition, the prediction of a total of 8 events is in agreement
with the observed $9\pm{3}$ data events. This further supports the
conclusion from the K-S test that we observe pion VCS events. 

From the limited statistics and sensitivity
of this first pion VCS experiment, we cannot determine the $\bar\alpha$
polarizability value, nor can we determine which value of $\bar\alpha$
is preferred. In a future experiment, the sensitivity to pion
polarizability may be increased~\cite{pvcs} by achieving a data set
in which the final $\gamma$ ($\pi$) has higher (lower)
energies. However, such data correspond to a lower cross section, and
therefore require a high luminosity experiment.   
\section{Conclusions}
The pion Virtual Compton Scattering via the reaction
$\pi{e}\rightarrow\pi{'}{e'}\gamma$ is observed. We developed and
implemented a simulation with a VCS event generator. We defined cuts that
allow background reduction and VCS signal extraction. The measured number
of reconstructed pion VCS events, and their distributions with respect
to the Mandelstam invariants, are in reasonable agreement with
theoretical expectations. 
The corresponding
$\pi$ VCS experimental cross section is 
$\sigma=38.8\pm{13}\,\mbox{nb}$, in
agreement with the theoretical expectation $\sigma=34.7\,\mbox{nb}$. 
\section{Acknowledgments}
The authors are indebted to the staffs of Fermi National Accelerator 
Laboratory, the Max--Planck--Institut f\"ur Kernphysik, Carnegie Mellon 
University, Petersburg Nuclear Physics Institute and Tel Aviv
University for invaluable technical support. We thank 
Drs. C. Unkmeir and S. Scherer for the VCS matrix element calculation. 

This project was supported in part by Bundesministerium f\"ur Bildung, 
Wissenschaft, Forschung und Technologie, Consejo Nacional de 
Ciencia y Tecnolog\'{\i}a {\nobreak (CONACyT)},
Conselho Nacional de Desenvolvimento Cient\'{\i}fico e Tecnol\'ogico,
Fondo de Apoyo a la Investigaci\'on (UASLP),
Funda\c{c}\~ao de Amparo \`a Pesquisa do Estado de S\~ao Paulo (FAPESP),
the Israel Science Foundation founded by the Israel Academy of Sciences
and 
Humanities, Istituto Nazionale de Fisica Nucleare (INFN),
the International Science Foundation (ISF),
the National Science Foundation (Phy \#9602178),
NATO (grant CR6.941058-1360/94),
the Russian Academy of Science,
the Russian Ministry of Science and Technology,
the Turkish Scientific and Technological Research Board (T\"{U}B\.ITAK),
the U.S. Department of Energy (DOE grant DE-FG02-91ER40664 and DOE
contract
number DE-AC02-76CHO3000), and
the U.S.-Israel Binational Science Foundation (BSF).
\section{Acknowledgments}
The authors are indebted to the staffs of Fermi National Accelerator 
Laboratory, the Max--Planck--Institut f\"ur Kernphysik, Carnegie Mellon 
University, Petersburg Nuclear Physics Institute and Tel Aviv
University for invaluable technical support. We thank 
Drs. C. Unkmeir and S. Scherer for the VCS matrix element calculation. 

This project was supported in part by Bundesministerium f\"ur Bildung, 
Wissenschaft, Forschung und Technologie, Consejo Nacional de 
Ciencia y Tecnolog\'{\i}a {\nobreak (CONACyT)},
Conselho Nacional de Desenvolvimento Cient\'{\i}fico e Tecnol\'ogico,
Fondo de Apoyo a la Investigaci\'on (UASLP),
Funda\c{c}\~ao de Amparo \`a Pesquisa do Estado de S\~ao Paulo (FAPESP),
the Israel Science Foundation founded by the Israel Academy of Sciences
and 
Humanities, Istituto Nazionale de Fisica Nucleare (INFN),
the International Science Foundation (ISF),
the National Science Foundation (Phy \#9602178),
NATO (grant CR6.941058-1360/94),
the Russian Academy of Science,
the Russian Ministry of Science and Technology,
the Turkish Scientific and Technological Research Board (T\"{U}B\.ITAK),
the U.S. Department of Energy (DOE grant DE-FG02-91ER40664 and DOE
contract
number DE-AC02-76CHO3000), and
the U.S.-Israel Binational Science Foundation (BSF).
 
\end{document}